\documentclass[reprint,aps,twocolumn,fixfloat,nofootinbib,superscriptaddress]{revtex4-2}
\usepackage{graphicx,amsmath,amsfonts,bm, color}

\usepackage{xcolor,hyperref}
\hypersetup{
   colorlinks,
   linkcolor={blue!50!black},
   citecolor={blue!50!black},
   urlcolor={blue!80!black}
}

\usepackage{enumitem}

\newcommand{\1}{^{\mbox{\tiny (1)}}}

\DeclareMathAlphabet{\mathitbf}{OML}{cmm}{b}{it}

\begin{document}

\title{Self-Driven Configurational Dynamics in Frustrated Spring-Mass Systems}

\author{Ori Saporta-Katz}
\affiliation{Computer Science and Applied Mathematics Department, Weizmann Institute of Science, Rehovot 7610001, Israel}
\author{Avraham Moriel}
\email{am7535@princeton.edu}
\altaffiliation[]{\\Current address: Department of Mechanical and Aerospace Engineering, Princeton University, New Jersey 08544, USA.}
\affiliation{Chemical and Biological Physics Department, Weizmann Institute of Science, Rehovot 7610001, Israel}

\begin{abstract}
Various physical systems relax mechanical frustration through configurational rearrangements. We examine such rearrangements via Hamiltonian dynamics of simple internally-stressed harmonic 4-mass systems. We demonstrate theoretically and numerically how mechanical frustration controls the underlying potential energy landscape. Then, we examine the harmonic 4-mass systems' Hamiltonian dynamics and relate the onset of chaotic motion to self-driven rearrangements. We show such configurational dynamics may occur without strong precursors, rendering such dynamics seemingly spontaneous.
\end{abstract}

\maketitle

\emph{Introduction.}---Mechanical frustration and internal stresses endow inanimate and active physical systems with unusual structural features. Internal stresses generate quasilocalized modes in amorphous materials and alter their thermodynamical properties~\cite{alexander1998amorphous,wang2005pressure,wyart2005rigidity,lerner2018frustration,mizuno2018phonon}. Mechanical frustration promotes nonextensive statistics and enables the emergence of complex patterns in spatially-extended systems~\cite{meiri2021cumulative,meiri2022cumulative,meiri2022bridging,han2008geometric,kang2014complex,aharoni2016internal}, and a wide range of conformations and structures for biomolecules and active matter ~\cite{nier2016inference,burla2019mechanical}. Understanding the role of mechanical frustration in generating these exotic properties is crucial.

Frustrated systems also exhibit diverse dynamical behaviors, including folding and conformational transitions, as means of stress relaxation~\cite{modes2012mechanical,zhang2014transformation,pierse2017distinguishing,milles2018calcium,tskhovrebova1997elasticity,guo2014mechanics,lieleg2011slow,zhang2021mechanical}. Such configurational dynamics are either triggered by external driving forces (e.g., mechanical loading) ~\cite{falk1998dynamics,lauridsen2002shear,cohen2004shear,dennin2004statistics,maloney2006amorphous,schall2007structural,lundberg2008reversible} or, are self-induced~\cite{hodge1995physical,phillips1996stretched,abou2001aging,mckenna2008diverging,feng2009residual,moreno2010stress,welch2013dynamics,qiao2016transition}. Athermal systems, in which external thermal fluctuations are insufficient to induce stress relaxations events, demonstrate that stress relaxation cannot be entirely attributed to thermally-activated or externally-driven processes~\cite{cipelletti2000universal,song2022microscopic,cipelletti2003universal,bouchaud2001anomalous,lahini2017nonmonotonic,lahini2023crackling,sollich1997rheology,balankin2011slow, shohat2023logarithmic}. Internal stresses clearly play a vital role in self-induced relaxation events~\cite {lieleg2011slow,gonzalez2021mechanical1,gonzalez2021mechanical2,lerner2017effect,kapteijns2021does}.

In this work, we use a mechanically frustrated, isolated, spring-mass system~\cite{moriel2021internally} to study the role of mechanical frustration on both structural characteristics and emerging dynamics, focusing on self-driven configurational rearrangements. We probe the systems' Hamiltonian dynamics, rendering all observed behaviors self-driven and not externally induced. We first study how frustration modifies the underlying potential energy landscape, the local and global energetic minima, and the transition state. These modifications affect the system's Hamiltonian dynamics and alter the onset of chaotic motion and configurational rearrangements. Finally, we show such configurational dynamics can occur without strong dynamic precursors. 

\begin{figure}[t]
\centering
\includegraphics[width=0.46\textwidth]{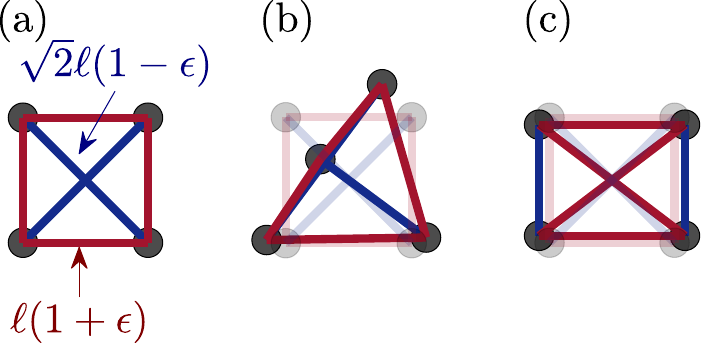}
\caption{(a) The square configuration, with the rest-lengths of the peripheral springs (red) and diagonal springs (blue) specified as a function of the amplitude of internal stress $\epsilon$. (b) The transition configuration between the square and rectangle stable mechanical minima. 
(c) The rectangle stable mechanical minima, where the two blue springs are now on the outer part. All panels are generated using $\epsilon\!=\!0.3$, and in panels (b,c), the square configuration is reproduced in the background for comparison.}
\label{fig:fig1}
\end{figure}

\emph{Frustrated spring-mass systems.}---To study internal stress's roles, we look for a simple spring-mass system supporting mechanical frustration. Previous works on the harmonic 3-mass systems demonstrated that such a simple system could self-induce rotations~\cite{katz2019self,katz2020regular}. As the $3$ masses have to align to allow a force-balanced internally-stressed state, we consider here a similar harmonic system composed of fully interacting $4$ particles (i.e., each particle has $3$ interactions) in $2$ dimensions~\cite{moriel2021internally}. The particles interact via geometrically-nonlinear springs of finite rest-length $\frac{k}{2}\left(r_{ij} - L_{ij}\right)^2$, where the spring-constant $k$ is set to unity, $r_{ij}\!\equiv\!\sqrt{(x_i-x_j)^2 + (y_i-y_j)^2}$ is the distance and $L_{ij}$ is the rest-length of the spring between the particles $i$ and $j$. The total potential energy $U$ is a sum over the pairwise interactions $\frac{k}{2}\sum_{\alpha=1}^6 \left(r_\alpha - L_\alpha\right)^2$ (where $\alpha$ runs over all pairs). Figure~\ref{fig:fig1}(a) shows a sketch of the system of interest.

The system is stress-free when each pairwise interaction exerts zero force. For a stress-free square configuration, $L_{ij}\!=\!\ell$ for peripheral pairs and $L_{ij}\!=\!\sqrt{2}\ell$ for pairs interacting along the diagonals ($\ell$ sets the length dimensions). If the net force over each particle vanishes, while the pairwise forces are \emph{non-zero}, the system is mechanically frustrated~\cite{alexander1998amorphous,moriel2021internally,mao2018maxwell,lubensky2015phonons}. The 4-mass system allows a frustrated state once the individual springs are \emph{not} at their rest lengths. We obtain such a state by changing the spring rest-lengths to $L_{ij}\!=\!\ell(1+\epsilon)$ for peripheral pairs, and $L_{ij}\!=\!\sqrt{2}\ell\left(1-\epsilon\right)$ along the diagonals~\cite{SM}, as shown in Fig.~\ref{fig:fig1}(a). We set $\ell$ to unity, such that the dimensionless $\epsilon$ captures the internal stress amplitude.

Taken together with the kinetic energy $K\!=\!\frac{1}{2 m} \sum_{i=1}^4 p_i^2$ of the 4-mass system ($i$ runs over the four particles, $p_i\!\equiv\!|\bm{p}_i|$, $\bm{p}_i\!=\!\left(p_i^x,p_i^y\right)$, and $m$ set to unity), the system's Hamiltonian $\mathcal{H}$ is~\cite{landau1976mechanics}
\begin{equation} \label{eq:Hamiltonian}
\mathcal{H} = K + U \ .
\end{equation}
The Hamiltonian dynamics of the system conserve the total energy, such that possible configurational changes are only self-induced (and are not externally triggered).

\begin{figure}[t]
\centering
\includegraphics[width=0.48\textwidth]{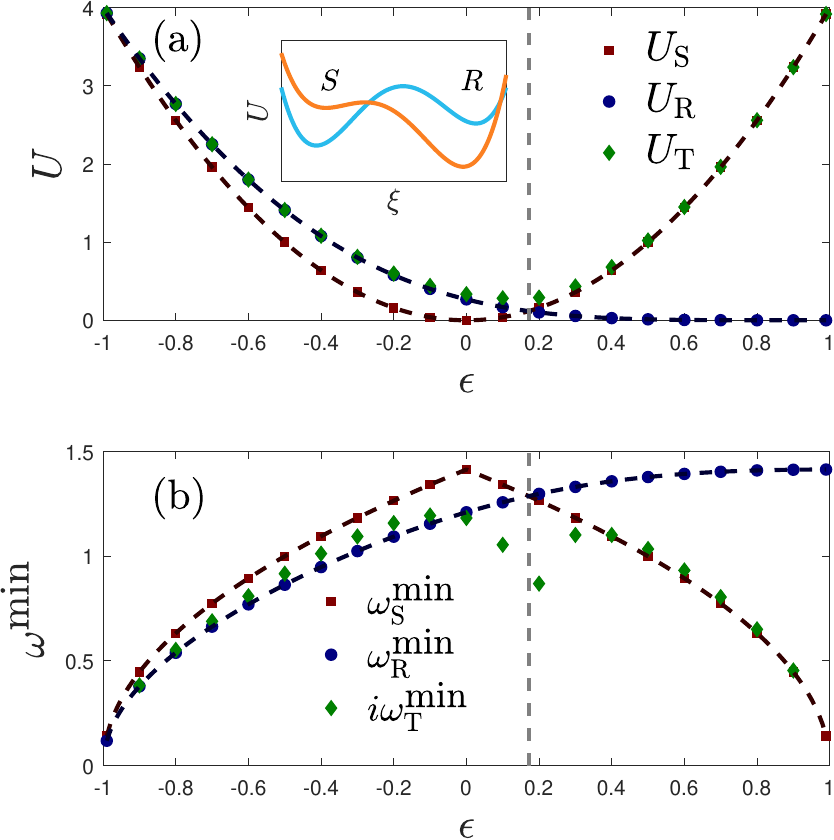}
\caption{(a) The potential energy values at the square configuration $U_{\rm{S}}$, the rectangular configuration $U_{\rm{R}}$, and the transition state $U_{\rm{T}}$. The $U_{\rm{S}}, U_{\rm{R}}$ data are obtained from random configuration sampling and numerical minimization of the potential energy. The $U_{\rm{T}}$ data are obtained via the Nudge Elastic Band algorithm between the square and rectangle configuration. Dashed lines indicate the theoretically predicted values~\cite{SM}, and the vertical dashed line denotes the critical $\epsilon_*$ value where the global minimum switches between the square geometry ($\epsilon\!<\!\epsilon_*$) to the rectangular configuration ($\epsilon\!>\!\epsilon_*$). Inset: a sketch of the potential energy landscape versus the reaction coordinate $\xi$, showing schematically that for $\epsilon\!<\!\epsilon_*$ the square is a global minimum (blue), and for $\epsilon\!>\!\epsilon_*$ the rectangular configuration becomes the global one (orange). (b) The minimal frequencies at the square and rectangular minima and the divergence rate at the transition configurations. $\omega^{\rm{min}}_{\rm{S}}$ and $\omega^{\rm{min}}_{\rm{R}}$ are associated with the minimal oscillation frequencies around the square and rectangle configurations, while for the transition state, $i\omega^{\rm{min}}_{\rm{T}}$ corresponds to the divergence rate from the transition state.}
\label{fig:fig2}
\end{figure}

\emph{Mechanical frustration modifies the energy landscape.}---We first focus on the implications of mechanical frustration on the underlying potential energy $U$. The potential energy at the square configuration $U_{\rm{S}}$ changes quadratically with $\epsilon$, as shown in Fig.~\ref{fig:fig2}(a). Expanding $U$ perturbatively around the square geometry, we obtain the Hessian $\bm{\mathcal{M}}\!\equiv\!\tfrac{\partial^2 U}{\partial \bm{r} \partial \bm{r}}$, its eigenmodes $\bm{\psi}$'s and their respective frequencies $\omega$'s via the eigenvalue equation $\bm{\mathcal{M}} \bm{\psi}_i \!=\! \omega_i^2 \bm{\psi}_i$ (without summation). The lowest frequency at the square configuration $\omega_{\rm{S}}^{\rm{min}}$ changes with $\epsilon$ as shown in Fig.~\ref{fig:fig2}(b), and becomes negative at $|\epsilon|\!=\!1$~\cite{moriel2021internally}, implying the square configuration becomes unstable. In what follows, we study the system in the stable range $|\epsilon|\!<\!1$.

While the square configuration is stable for $|\epsilon|\!<\!1$, it does not imply that it is the \emph{only} energetic minimum in the entire energy landscape. A particle swap along the two diagonals will result in a geometrically-identical square configuration; swapping any peripheral pair will yield a rectangular shape with potential energy $U_{\rm{R}}$ and minimal frequency $\omega_{\rm{R}}^{\rm{min}}$ as sketched in Fig.~\ref{fig:fig1}(c). This configuration is potentially another energetic minimum~\cite{SM}. We numerically test and verify that the rectangular configuration serves as an additional minimum by initiating the 4 masses at random configurations and minimizing their potential energy to recover the closest minimum. The results obtained from this numerical procedure exactly match our analytical calculations, as shown in Fig.~\ref{fig:fig2}(a). 

For $\epsilon$ values below a critical value~\cite{SM} $\epsilon_*\!=\!3-2\sqrt{2}$ the square geometry serves as a \emph{global} minimum of the potential energy landscape, while for $\epsilon\!>\!\epsilon_*$ the rectangle configuration becomes the \emph{global} minimum, and the initial square geometry turns into a \emph{local} minimum [interestingly $\omega_{\rm{R}}^{\rm{min}}$ crosses $\omega_{\rm{S}}^{\rm{min}}$ at $\epsilon_*$, shown in Fig.~\ref{fig:fig2}(b)]. The value of $\epsilon_*$ is plotted in a dashed vertical line in Fig.~\ref{fig:fig2}(a), and we sketch this scenario in the inset of Fig.~\ref{fig:fig2}(a).

The existence of several energetic minima implies transition states exist in between. The transition state between the square and rectangular geometry resembles the one sketched in Fig.~\ref{fig:fig1}(b), where the top two particles swap. To detect and probe the transition state between the two minima, we employ the Nudge Elastic Band algorithm~\cite{tadmor2011modeling,wales2003energy} followed by force amplitude minimization. We plot the potential energy at the transition state $U_{\rm{T}}$ in Fig.~\ref{fig:fig2}(a). For $\epsilon\!<\!\epsilon_*$ the transition potential energy is well-approximated by $U_{\rm{R}}$, while for $\epsilon\!>\!\epsilon_*$ it is closer to $U_{\rm{S}}$ (though $U_{\rm{T}} > U_{\rm{S}}, U_{\rm{R}}$). 

At the transition state, the smallest eigenvalue is negative by definition~\cite{strogatz2018nonlinear}. We plotted the associated frequency $i \omega^{\rm{min}}_{\rm{T}}$ in Fig.~\ref{fig:fig2}(b). Unlike the stable mechanical minima, this value corresponds to a divergence rate, not an oscillation frequency. Once the system is provided with sufficient energy, it can pass between the two minima. The internal stress level changes the potential energy minima (local and global) and the potential energy barrier. Thus, we hypothesize mechanical frustration will also play a crucial role in the Hamiltonian dynamics of such a system, specifically in the onset of chaotic motion and configurational rearrangements.

\begin{figure}[b]
\centering
\includegraphics[width=0.48\textwidth]{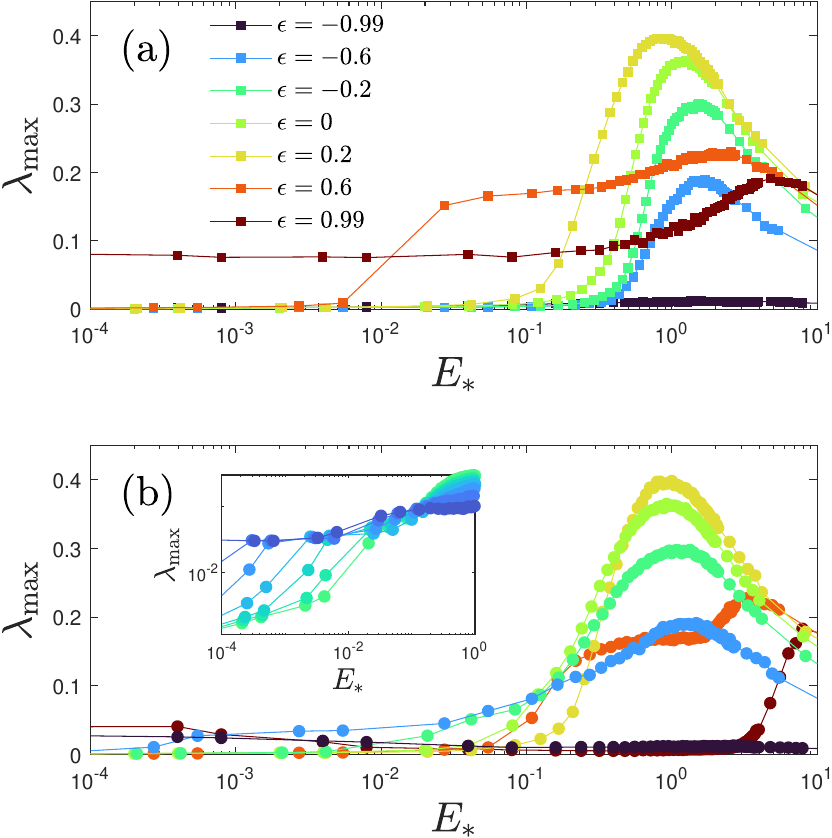}
\caption{(a) The maximal Lyapunov exponents $\lambda_{\rm{max}}$ for trajectories initiated with the square geometry, for different $\epsilon$ values and wide excess energy $E_*$ ranges (each point corresponds to ensemble average over $100$ realizations). Above $\epsilon_*$, $\lambda_{\rm{max}}$ plateaus to a finite value at low $E_*$. (b) $\lambda_{\rm{max}}$ for trajectories initiated with the rectangle geometry, for $\epsilon$ values and energy range as in panel (a) (each point corresponds to ensemble average over $100$ realizations). A $\lambda_{\rm{max}}$ plateau occurs at $\epsilon \!<\!\epsilon_*$ values, where the rectangular geometry is a local minimum of the potential energy landscape. Inset: a log-log plot zoomed-in on the plateau at low $E_*$ values for $\epsilon$ between $-0.8$ (blue) and $-0.2$ (green).}
\label{fig:fig3}
\end{figure}

\emph{Hamiltonian dynamics and chaotic motion.}---What special dynamical behaviors emerge due to mechanical frustration? To probe the intrinsic dynamics of the frustrated spring-mass systems, we consider the Hamiltonian dynamics governed by Eq.~\eqref{eq:Hamiltonian}. We initialize the system at its internally-stressed, square configuration and set the initial velocities randomly. We remove any linear and angular momentum from the initial velocities and ensure the total excess kinetic energy provided to the system is $E_*$~\cite{SM}. Overall, the initial configuration corresponds to a potential energy $U_{\rm{S}}(\epsilon)$, and an initial kinetic energy $K\!=\!E_*$ (with no linear and angular momentum). We probe the self-driven dynamics via measuring the maximal Lyapunov exponent $\lambda_{\rm{max}}$~\cite{strogatz2018nonlinear,ramasubramanian2000comparative} numerically, at the end of our simulations~\cite{wolf1985determining,Govorukhin2023Calculation,SM}. We plot $\lambda_{\rm{max}}$ averaged over an ensemble of 100 realizations in Fig.~\ref{fig:fig3}(a).

At low $E_*$ values, the system oscillates around the square configuration (as it is a stable mechanical minimum~\cite{landau1976mechanics,strogatz2018nonlinear}, according to the KAM theory~\cite{lichtenberg2013regular}), and $\lambda_{\rm{max}}$ is expected to be low~\cite{katz2020regular,strogatz2018nonlinear}. At high $E_*$, the values of $L_{ij}$ become irrelevant, and the system is expected to exhibit regular dynamics again~\cite{katz2020regular}. For intermediate excess energies, $\lambda_{\rm{max}}$ peaks approximately at the energy needed to compress a single bond (varying with $\epsilon$). These results are qualitatively similar to those of the harmonic 3-mass system~\cite{katz2020regular}.

Surprisingly, at low energies and for $\epsilon\!>\!\epsilon_*$, $\lambda_{\rm{max}}$ plateaus at energy scales lower than those associated with a single bond compression, as shown in Fig.~\ref{fig:fig3}(a). As $\epsilon\!\rightarrow\!1$ the plateau persists to lower energy scales. This peculiar behavior has not been observed in the case of a triangle~\cite{katz2020regular} and seems to emerge specifically due to mechanical frustration. Intrigued by this phenomenon, we repeat the same procedure, now initializing the system at the rectangular configuration. This results in $\lambda_{\rm{max}}$ shown in Fig.~\ref{fig:fig3}(b). Now, we observe a plateau for sufficiently low $\epsilon$ values (we discuss the appearance of plateaus at higher $\epsilon$ values in~\cite{SM}). The $\lambda_{\rm{max}}$ plateaus seem to correlate with the global-to-local transition of the two configurations. We hypothesize the transition state governs this \emph{dynamical} observable.

At low energies, the system's trajectories occupy phase space regions close to the mechanically stable states. Such trajectories describe the system's oscillations around the stable mechanical minimum. At some stage, $E_*$ is sufficient for the system to pass over the transition state, resulting in configurational rearrangement; the available phase space changes dramatically as it now includes several stable minima and saddles. We hypothesize the saddles act as effective scatterers of the trajectories, causing them to diverge from one another, increasing the Lyapunov exponent~\cite{strogatz2018nonlinear}. Further increasing $E_*$, the phase space volume includes more saddles and minima, possibly scattering the trajectories even more strongly. Eventually, all minima and saddles are included, and the phase-space volume increases trivially with increased available energy. 

\begin{figure}[t]
\centering
\includegraphics[width=0.48\textwidth]{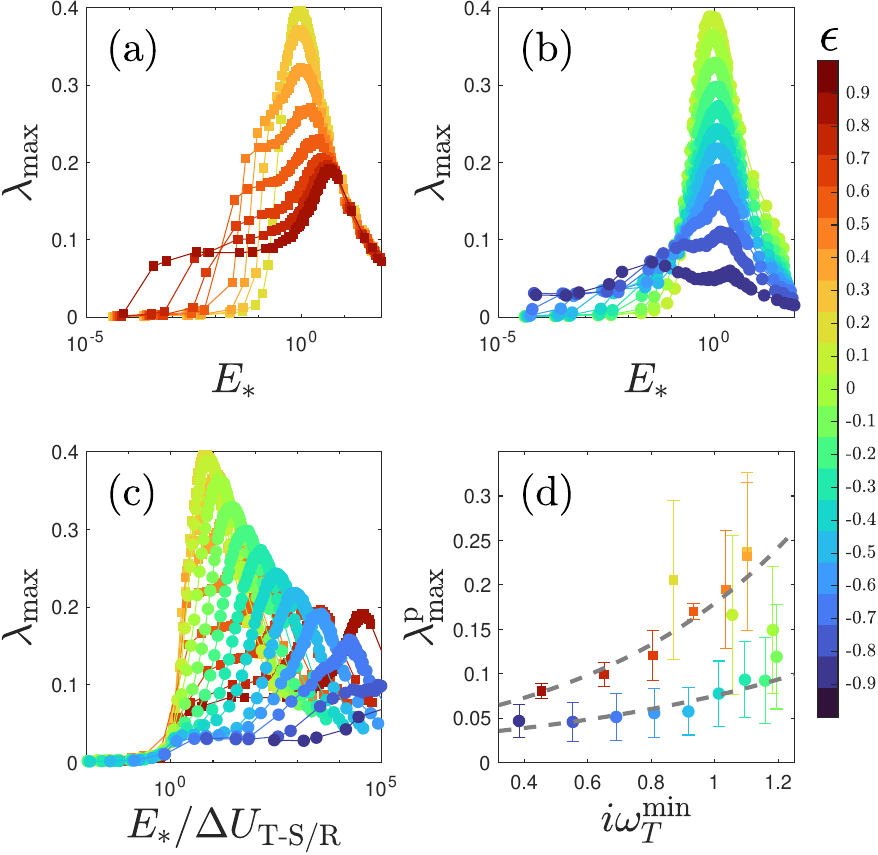}
\caption{(a) $\lambda_{\rm{max}}$ from the square trajectories (with $\epsilon \!>\! \epsilon_*$) versus $E_*$. (b) $\lambda_{\rm{max}}$ form the rectangular trajectories (with $\epsilon \!<\! \epsilon_*$) versus $E_*$. (c) The same $\lambda_{\rm{max}}$ from panels (a) and (b) [$\Box$ corresponds to the square data of panel (a) and $\circ$ corresponds to the rectangular data from panel (b)], versus $E_*$ rescaled by the barrier energy $\Delta U_{\rm{T-S}}$ for data from panel (a), and $\Delta U_{\rm{T-R}}$ for data from panel (b). (d) The approximated plateau value of the Lyapunov exponent $\lambda_{\rm{max}}^p$ (an average over the first $7$ points above the barrier energy) versus $i\omega_{\rm{T}}^{\rm{min}}$ (error bars indicate the standard deviation of the $7$ points). For well-approximated plateaus (small errorbars), $\lambda_{\rm{max}}^p$ forms a function of $i\omega_{T}^{\rm{min}}$ (the two gray exponential curves serve as a guide to the eye). $\Box$ corresponds to the square data and $\circ$ corresponds to the rectangular data. The color bar shows the $\epsilon$ values used in the $4$ panels.}
\label{fig:fig4}
\end{figure}

To test this argument, we focus on the transition between the square and rectangular configurations. We consider trajectories with $\epsilon$ values in which the initial configuration is the \emph{local} minimum, as shown in Fig.~\ref{fig:fig4}(a)-(b). Once we rescale $E_*$ by the relative barrier energy $\Delta U_{\rm{T-S}} \!\equiv\! U_{\rm{T}} - U_{\rm{S}}$ ($\Delta U_{\rm{T-R}} \!\equiv\! U_{\rm{T}} - U_{\rm{R}}$) for the square (rectangular) trajectories, the initial increase in $\lambda_{\rm{max}}$ collapses, as shown in Fig.~\ref{fig:fig4}(c). Then, we approximate the value of $\lambda_{\rm{max}}$ at the plateau, $\lambda_{\rm{max}}^p$,  by averaging over the first $7$ entries above the barrier energy. We plot the resulting $\lambda_{\rm{max}}^p$ versus $i\omega_{\rm{min}}^{\rm{T}}$ in Fig.~\ref{fig:fig4}(d), demonstrating that $\lambda_{\rm{max}}^p\left(i\omega_{\rm{min}}^{\rm{T}}\right)$ for well-detected plateaus~\cite{strogatz2018nonlinear}. The results of Fig.~\ref{fig:fig4}(c)-(d) demonstrate the relation between the transition state --- a \emph{potential energy landscape} property ---, and an increased Lyapunov exponent --- a \emph{dynamical} observable~\cite{wu2005nonlinearity,huang2007dynamical,chao2009dynamical}.

\emph{Configurational dynamics.}---Next, we follow trajectories originating from the square configuration for $\epsilon\!>\!\epsilon_*$, with $E_*$ slightly above $ \Delta U_{\rm{T} - \rm{S}}$ to detect self-driven configurational rearrangements. The system oscillates around the initial configuration until it passes over the energetic barrier, releasing its stored potential energy. These dynamics correspond to configurational rearrangements like that shown in Fig.~\ref{fig:fig1}.

\begin{figure}[t!]
\centering
\includegraphics[width=0.49\textwidth]{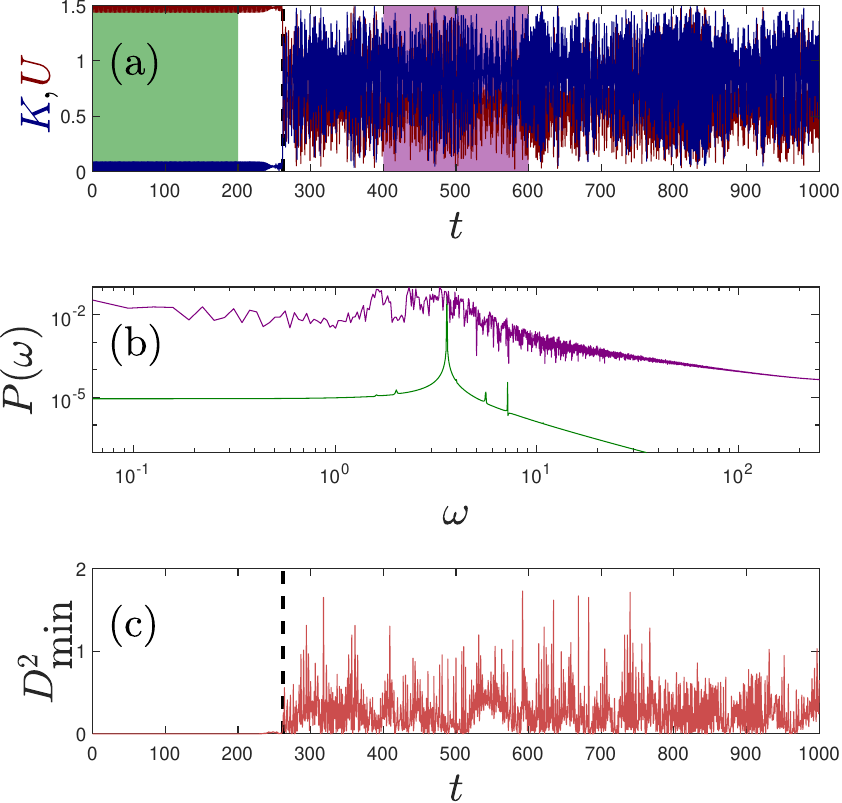}
\caption{(a) Kinetic (blue) and potential (red) energies as a function of time for $\epsilon\!=\!0.3$ and $E_*\!\simeq\!0.08$ ($\Delta U_{\rm{T-S}} \!\simeq\!0.07$). A vertical black dashed line marks the first crossing between $U$ and $K$. Early-time and late-time regions are marked with green- and purple-shaded regions, respectively. (b) The power spectrum of the early-time and late-time shaded regions of (a). The early-time power spectrum of $U$ includes discrete peaks representative of regular dynamics, while the late-time power spectrum contains contributions at all frequencies. (c) $D^2_{\rm{min}}$ measure of non-affine motion as a function of time for the same trajectory. The moment of the first crossing is marked in a vertical black dashed line, indicating the crossing of $U$ and $K$ is accompanied by a large non-affine motion.}
\label{fig:fig5}
\end{figure}

Are there any precursors to these rearrangements? We plot various dynamic measurements from an example trajectory in Fig.~\ref{fig:fig5}. Fig.~\ref{fig:fig5}(a) shows $K$ and $U$ throughout the dynamics, revealing the finite-time release of the initially stored potential energy, as indicated by the vertical dashed line. This release signifies passing away from the square energy basin to other regions in phase space. The system oscillates uniformly until it passes over the energetic barrier, after which the dynamics become irregular. 

To further examine the dynamics before and after the barrier crossing, we analyze the power spectrum of $U$ in the early and late stages of the simulations in Fig.~\ref{fig:fig5}(b) [corresponding to the shaded regions in Fig.~\ref{fig:fig5}(a)]. The early-time power spectrum shows discrete peaks and does not hint at the impending configurational change. After crossing the energetic barrier, various frequencies fill the power spectrum, indicating the trajectory becomes irregular. 

We also measured the degree of non-affine motion via the $D^2_{\rm{min}}$ measure~\cite{falk1998dynamics} in Fig.~\ref{fig:fig5}(c). The $D^2_{\rm{min}}$ measure peaks at the first crossing of $U$ and $K$, indicating that passing from the square energy basin to the outer phase-space is associated with a non-affine deformation of the system, suggesting the system underwent a dramatic configurational rearrangement to pass over the barrier. While other dynamical behaviors are possible~\cite{SM}, we haven't observed strong dynamical precursors to such configurational changes, rendering these self-driven events ``spontaneous''~\cite{lieleg2011slow}. 

\emph{Discussion.}---We studied self-driven configurational dynamics in frustrated spring 4-mass systems. Changing the internal stress amplitude $\epsilon$ varied the springs' rest lengths and modified the stable mechanical minima and the transition states between them. These modifications yielded unique Lyapunov exponent plateaus. The plateaus arise at energies comparable to the barrier energies $U_{\rm{T}} - U_{\rm{S/R}}$ and are affected by the eigenvalue associated with the saddle's most unstable direction. Finally, we have demonstrated how trajectories with sufficient excess energy could seem completely regular before undergoing configurational changes and overcoming the energetic barrier. 

The isolated spring-mass systems considered above allowed us to vary the internal stress continuously and study the emerging self-driven dynamics. This systematic variation allowed us to relate the emerging plateau in $\lambda_{\rm{max}}^p$ with $i\omega_{T}^{\rm{min}}$, which is unavailable once a specific molecule or material is considered~\cite{wu2005nonlinearity,huang2007dynamical,chao2009dynamical}. Mathematically, this demonstrates the effects of saddle points on $\lambda_{\rm{max}}$ when the underlying energy landscape is complex (e.g., mixed systems). 

The isolated system considered above is not spatially extended, preventing us from further exploring the spatial signatures of such configurational dynamics~\cite{cipelletti2000universal,cipelletti2003universal,lieleg2011slow}. Spatially extended systems undergoing configurational rearrangements can divert and spread the released energy to other system parts or an external bath, possibly inducing avalanches~\cite{lauridsen2002shear,song2022microscopic,shohat2023logarithmic}. The isolated 4-mass system cannot support this behavior. We suspect embedding a mechanically frustrated element within a stress-free spatially extended system will suppress the configurational dynamics observed above (as these will be more energetically costly). Including positional disordered, internally stressed elements may lower the energetic barriers and enable studying the spatiotemporal dynamics triggered by self-induced configurational dynamics. 


We thank Yuri Lubomirsky for his insightful comments and discussions and Dor Shohat, Yoav Lahini, and Eran Bouchbinder for commenting on the manuscript. O.S.K. acknowledges the support of a research grant from the Yotam Project and the
Weizmann Institute Sustainability and Energy Research Initiative; and the support of the Séphora Berrebi
Scholarship in Mathematics. A.M. acknowledges support from the Minerva Foundation with funding from the Federal German Ministry for Education and Research, the Ben May Center for Chemical Theory and Computation, and the Harold Perlman Family.

O.S.K. and A.M. designed the research. A.M. performed numerical simulations. O.S.K. and A.M. performed theoretical calculations and wrote the manuscript.



\onecolumngrid
\newpage
\begin{center}
\textbf{\large Supplemental Materials for: \\ ``Self-Driven Configurational Dynamics in Frustrated Spring-Mass Systems''}
\end{center}
\twocolumngrid
\setcounter{equation}{0}
\setcounter{figure}{0}
\setcounter{table}{0}
\setcounter{section}{0}
\setcounter{page}{1}
\makeatletter
\renewcommand{\theequation}{S\arabic{equation}}
\renewcommand{\thesection}{S-\Roman{section}}
\renewcommand{\thefigure}{S\arabic{figure}}
\renewcommand*{\thepage}{S\arabic{page}}
\renewcommand{\bibnumfmt}[1]{[S#1]}
\renewcommand{\citenumfont}[1]{S#1}

In this Supplemental Materials, we provide the procedure to obtain the interaction lengths and stiffnesses from the minimal complex construction, a schematic derivation for analytically obtaining the rectangular energetic minimum, and explicit expressions for the potential energy values at the square and rectangle configurations. We also provide details regarding the initialization and integration of the equations of motion and the Lyapunov exponent calculations. We then briefly discuss the additional plateaus observed in Fig.3(b) in the manuscript. Finally, we provide additional examples of the dynamics observed slightly above the barrier energy.

\subsection{From minimal complexes to springs constants and rest lengths}
We start from the minimal complex construction proposed in~\cite{Smoriel2021internally}, with $\varphi_\alpha$ being the pairwise potential of the $\alpha$th interaction. The Hessian $\bm{\mathcal{M}}$ contains two contributions~\cite{Slerner2018frustration} --- one coming from the elastic stiffness $\varphi_\alpha'' \!=\! \partial^2 \varphi_\alpha / \partial r_\alpha \partial r_\alpha$ (with $r_\alpha\!\equiv\!|\bm{r}_i-\bm{r}_j|$ being the magnitude of the difference vector between the interacting particles $i$ and $j$ in the $\alpha$th bond), and one from the internal stresses $\varphi_\alpha' \!=\! \partial \varphi_\alpha / \partial r_\alpha \!\equiv\!- f_\alpha$. In~\cite{Smoriel2021internally}, the internal stresses were obtained by the null-space vector, yielding zero net force on the particles. As we initially consider the square geometry, this null-space vector yields the values $f_\alpha \!=\! \epsilon$ for $\alpha$ of the peripheral interactions, and $f_\alpha \!=\! -\sqrt{2} \epsilon$ for the diagonal interactions. 

Now, we assume the particles interact via geometrically nonlinear springs. The pairwise interaction potential takes the form $\varphi_\alpha \!=\! \frac{k_\alpha}{2}\left(r_\alpha - L_\alpha\right)^2$. We choose $k_\alpha$ and $L_\alpha$ such that $\varphi_\alpha''\!=\!1$ and $\varphi_\alpha' \!=\! -f_\alpha$ at the square configuration, to emphasize the role of mechanical frustration (e.g., as opposed to the elastic stiffnesses). Setting $k_\alpha \!=\! 1$ satisfies the first condition. The second condition is satisfied by choosing $L_\alpha \!=\! r_\alpha - (\varphi_\alpha' /  \varphi_\alpha'')$, where $r_\alpha$ is the distance of the $\alpha$th pair in the rest configuration (i.e., either $\ell$ for peripheral pairs or $\sqrt{2}\ell$ for diagonal pairs). As mentioned in the main text, we set $\ell$ to unity.

\subsection{Analytical derivation of the rectangular minima and potential energy values}
We hypothesize an exchange along a single bond potentially yields an additional minimum. In the case of an exchange along either of the diagonals, the resulting shape is geometrically identical to the initial square configuration. Switching along peripheral interactions could render a geometrically different configuration as a minimum.

To study this possibility, we note that overall, there are $8$ configurational degrees of freedom. However, translational and rotational invariance implies $3$ of these are unnecessary, and the positions of the particles could be expressed using the remaining $5$ degrees of freedom. To study the rectangular configuration, we first use the translational invariance to fix the $x$ and $y$ coordinates of the bottom left particle to the origin. We then use rotational invariance to fix the bottom right particle to the $x$ axis. Finally, we parameterize the rectangular configuration by its diagonal length $\tilde{L}$ and the angle between the diagonal and the $x$ axis $\phi$. If, in the square geometry, we considered particles at $(0,0)$, $(1,0)$, $(1,1)$ and $(0,1)$, we now parameterize the positions of the particles at the rectangular state as $(0,0)$, $(\tilde{L} \cos \phi,0)$, $(0,\tilde{L} \sin \phi)$ and $(\tilde{L} \cos \phi,\tilde{L} \sin \phi)$ (with particles $3$ and $4$ exchanged). 

If a state produces zero net forces, it is potentially an additional minimum. Expressing the potential energy $U$ using the prescribed positions and solving for $\partial U / \partial \bm{r} \!=\! \bm{0} $~\cite{SMathematica}, we obtain a solution of the form $\tilde{L} \!=\!\frac{1}{2}\left(1 + \epsilon + \sqrt{3 + \epsilon\left(3 \epsilon - 2\right)}\right)$, and $\phi\!=\! \arctan \left[(\sqrt{2} \left(1-\epsilon\right))/(1+\epsilon)\right]$. We also calculate the potential energy associated with the square and rectangle configuration by plugging in the particles' positions in the potential energy $U$ to obtain $U_{\rm{S}} = 4 \epsilon^2$ and $U_{\rm{R}} \!=\! \frac{1}{2}\left(1 + \epsilon -\sqrt{3+\epsilon\left(3\epsilon - 2\right)}\right)^2$. At $\epsilon\!=\!-1$ and at $\epsilon_*\!=\!3-2\sqrt{2}$,  $U_{\rm{R}}\!=\!U_{\rm{S}}$.

These energetic minima are degenerate, as we may choose any of the $4$ peripheral interactions to perform this exchange. Additionally, the paths between the square and a rectangular minimum are also degenerate --- of the two particles exchanging places, one could go above or below the other. We suspect $\lambda_{\rm{max}}^p$ may depend on these degeneracies. We also expect some of these degeneracies to disappear in the presence of positional disorder~\cite{Smoriel2021internally}.

\subsection{Simulation details and Lyapunov exponents}
As mentioned in the manuscript, in all our simulations, the $4$ particle system was initiated at the perfect square or rectangular configurations (see the explicit positions above). The initial velocities/momenta $\bm{p}^0$ are picked randomly from a uniform distribution (recall all masses are set to unity $m\!=\!1$). We then use a projection operator to remove any overall translations and rotations.

To construct the projection operator, we first find three vectors that represent (1) linear momentum in the $x$ direction, (2) linear momentum in the $y$ direction, and (3) angular momentum. The vector $\bm{j}_x$ that induces center of mass velocity in the $x$ direction takes a value of unity in the $x$ components and $0$ in the $y$ components, represented here as $\bm{j}_x \!=\! \left(1,0,1,0,1,0,1,0\right)$. The vector $\bm{j}_y$ is constructed in a similar fashion as $\bm{j}_y \!=\!\left(0,1,0,1,0,1,0,1\right)$. The vector representing rotation is $\bm{j}_a \!=\! \left(r_1^y - r_{\rm{cm}}^y, r_{\rm{cm}}^x - r_1^x,...,r_4^y - r_{\rm{cm}}^y, r_{\rm{cm}}^x - r_4^x\right)$.

We then stack these vectors in an $3$-by-$8$ matrix as $\bm{\mathcal{J}}\!\equiv\!\left[\bm{j}_x;\bm{j}_y;\bm{j}_a\right]$ (here we use $;$ to denote a new row). The square ($8$-by$8$) projection matrix $\mathcal{P}\!\equiv\! \bm{\mathcal{J}}^T\left(\bm{\mathcal{J}}\bm{\mathcal{J}}^T\right)^{-1}\bm{\mathcal{J}}$ will extract the center of mass translation and rotation from the initial momenta~\cite{Sstrang2022introduction}. We use $\tilde{\bm{p}}^0\!\equiv\!\bm{p}^0\left(\mathcal{I} - \mathcal{P}\right)$ to obtain initial momenta without these components, $\tilde{\bm{p}}^0$. To ensure the system gets exactly $E_*$ excess energy, we normalize $\left|\tilde{\bm{p}}^0\right|\!=\!\sqrt{2 E_*}$. Hamilton's equations of motion~\cite{Slandau1976mechanics} take the form $\dot{\bm{r}}_i = \frac{\partial \mathcal{H}}{\partial \bm{p}_i}$, $\dot{\bm{p}}_i = -\frac{\partial \mathcal{H}}{\partial \bm{r}_i}$, where the Hamiltonian $\mathcal{H}$ is specified in the manuscript, and the initial conditions specified above. In our simulations, we measure time in units of $\tau_0\!=\!\sqrt{m/k}$.

We integrate the dynamics in time using one of two methods. To obtain the Lyapunov exponents, we use~\cite{SGovorukhin2023Calculation}, integrating the equations of motion via MATLAB's ODE45~\cite{SMATLAB}. As the initial conditions are random, we average over 100 independent runs for each $\epsilon, E_*$ combination. The Lyapunov exponents are measured at the end of the simulation, allowing the exponents to converge. When the Lyapunov exponents are not of interest (e.g., in Fig.5 in the manuscript), we use an alternative, modified symplectic integrator~\cite{SSymplectic} (where we adapted a Leapfrog algorithm). For both algorithms, we use $dt\!=\!0.01\tau_0$, and run the dynamics for $1000 \tau_0$.

\begin{figure}[t]
\centering
\includegraphics[width=0.48\textwidth]{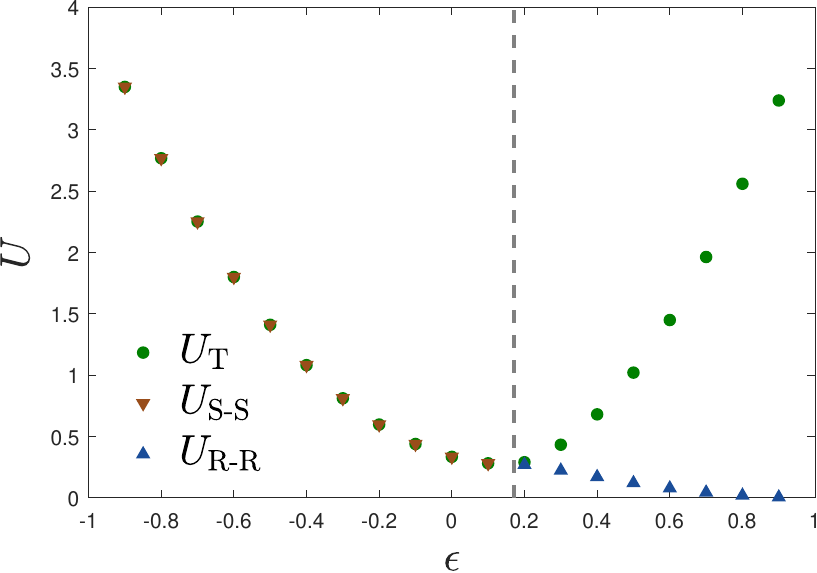}
\caption{The transition state energy $U_{\rm{T}}$ between the square and rectangle configurations, together with the energy barriers between two square configurations $\Delta U_{\rm{S-S}}$, and the energy barrier between two rectangular configurations $\Delta U_{\rm{R-R}}$ for a wide range of $\epsilon$ values. Below $\epsilon_*$, the local to global barrier $\Delta U_{\rm{T}}$ is similar to $\Delta U_{\rm{S-S}}$. A system initialized at the square geometry could hop between these states once $E_*\!>\! \Delta U_{\rm{S-S}}, \Delta U_{\rm{T}}$. For $\epsilon \!>\!\epsilon_*$, $\Delta U_{\rm{R-R}} \! \ll \Delta U_{\rm{T}}$, implying a rectangular system could rearrange to other rectangular configurations at energies much lower than that required for it to rearrange to a square configuration.}
\label{fig:FigS1}
\end{figure}

\subsection{Additional plateaus and scaling}
The plateaus in Fig.3(b) appear not only for low $\epsilon$ values but also for high values of $\epsilon$ [e.g., the $\epsilon\!=\!0.6$ curve in Fig.3(b)]. As mentioned in the manuscript, the appearance of plateaus in $\lambda_{\rm{max}}$ suggests an available transition state. However, the transition between the square and rectangular configuration is unavailable here, as it requires more energy than $E_*$ (see Fig.2 in the manuscript). 

\begin{figure*}[ht!]
\centering
\includegraphics[width=\textwidth]{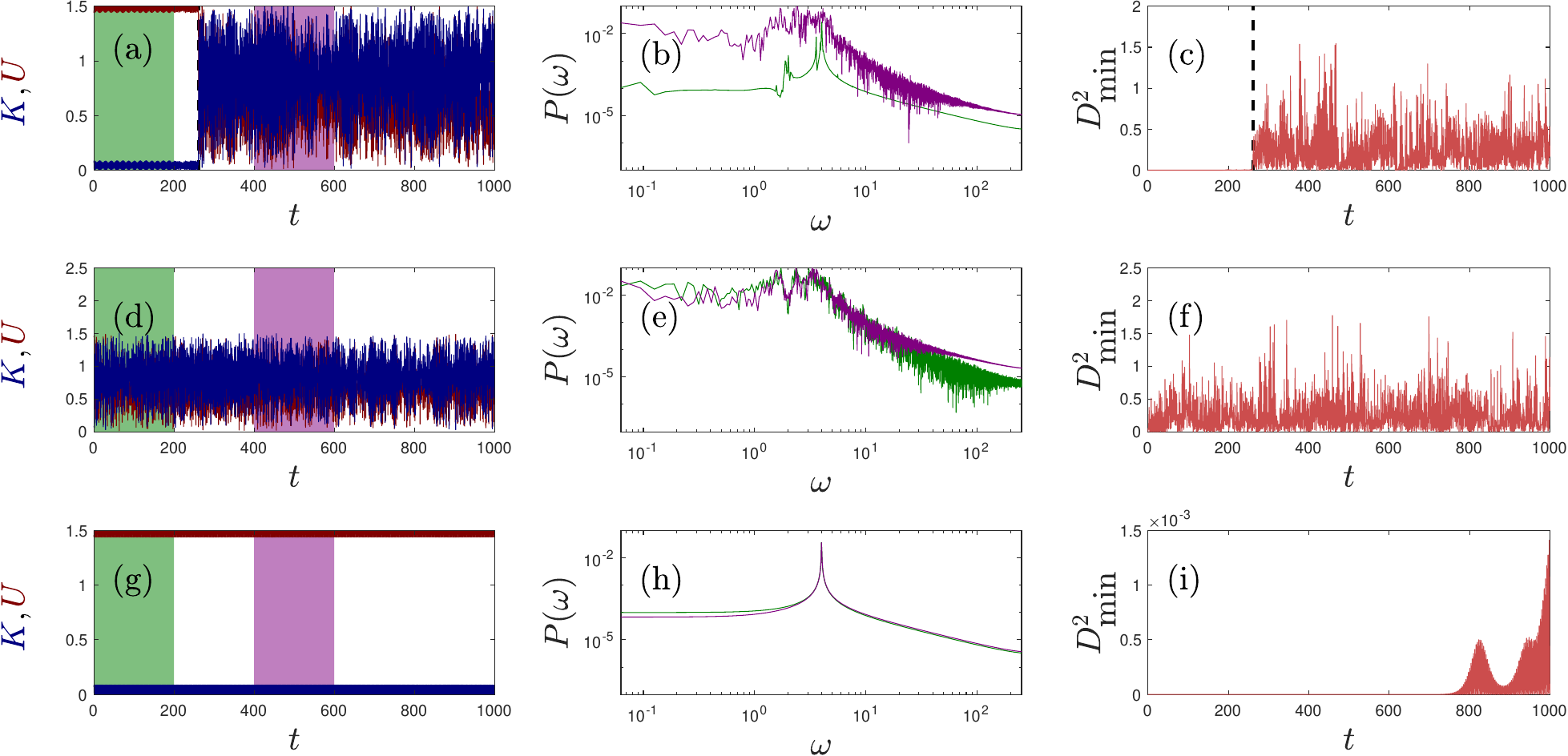}
\caption{(a) Kinetic (blue) and potential (red) energies as a function of time for $\epsilon\!=\!0.3$ and $E_*\!\simeq\!0.08$ ($\Delta U_{\rm{T-S}} \!\simeq\!0.07$). The first crossing between $U$ and $K$ is marked in a vertical black dashed line. Early-time and late-time regions are marked with green- and purple-shaded regions, respectively. (b) The power spectrum of the early-time and late-time shaded regions of (a). The early-time power spectrum already hints that the trajectory is chaotic, showing multiple frequencies, while the late-time power spectrum includes high-frequency contributions absent in the early-time power spectrum. (c) $D^2_{\rm{min}}$ measure of non-affine motion as a function of time for the same trajectory. The moment of the first crossing (i.e., the same one as in the top panel) is marked in a vertical black dashed line, indicating the crossing of the potential and kinetic energies is accompanied by large non-affine motion. (d) $U$ and $K$ for a trajectory with the same $\epsilon$ and $E_*$ as in the top row, but with different initial velocities. The system passes over the transition state very early on. (e) The early- and late-times power spectra of (d). The two power spectra show similar levels of noise. (f) The $D^2_{\rm{min}}$ measure of non-affine motion as a function of time, demonstrating the trajectory undergoes non-affine motion throughout the dynamics. A trajectory with the same $\epsilon$ and $E_*$, with different initial velocities. Here, the initial conditions yield regular dynamics over the entire trajectory. (h) The early- and late-times power spectra of (g) show a single frequency of oscillations (in both early and late times). (i) The $D^2_{\rm{min}}$ measure of non-affine motion showing the trajectory performs affine motion throughout the simulation.}
\label{fig:FigS2}
\end{figure*}

This suggests additional transition states become available to the system at lower energy values. We suspect these transition states are associated with exchanges that result in geometrically-identical configurations. For example, we suspect the plateau occurring in Fig.3(b) for $\epsilon\!=\!0.6$ at $E_*\!\simeq\!0.1$ results from the transition state between two rectangular configurations.

Revisiting Fig.1(a) in the manuscript, we notice that an exchange of two particles interacting via the diagonal (blue) interactions does not result in a geometrically different configuration. Interchange along any peripheral (red) interaction results in the configurational change shown in Fig.1. The same applies to transformations starting from the rectangular configuration shown in Fig.1(c) --- an exchange along the blue interactions results in a similar geometry, while an exchange along the red interactions could cause a configurational change (e.g., an exchange along the top red interaction results in a square configuration, while an interchange of the diagonal red interaction will result in a similar, though rotated, rectangular geometry).  

This observation suggests there are always several energy scales --- some associated with transformation into different configurations, while others associated with transformation into similar geometrical structures. For the red interactions, the energy scale associated with an exchange scales with $\ell (1+\epsilon)$, while the energy scale associated with an exchange across the blue interactions is $\sqrt{2}\ell(1-\epsilon)$. For $\epsilon \!<\!\epsilon_*$ an exchange along the red interactions is energetically favorable, regardless of the initial configuration. For $\epsilon \!>\! \epsilon_*$, an exchange across the blue interactions becomes less energetically demanding than an exchange along the red interactions. 

This simple analysis has major implications regarding the observed configurational dynamics. It first implies that, for a system initiated at the rectangular local energetic minimum for $\epsilon \!<\! \epsilon_*$, the first possible transition will involve a configurational change --- from the rectangular to the square configuration. It also implies a system initialized at the square configuration will exchange peripheral pairs easily. If we now consider a square-to-square rearrangement, the scaling analysis implies that for $\epsilon \!<\! \epsilon_*$ there would be two peripheral exchanges and not a single diagonal exchange. This means the barrier between two square configurations should be comparable to that of a square-rectangle rearrangement for the $\epsilon \!<\! \epsilon_*$ regime, explaining the absence of plateaus for $E_* \!<\! \Delta U_{\rm{T}-\rm{S}}$.

Above $\epsilon_*$, the square geometry is the local energetic minimum. A system initialized at this configuration will rearrange to the rectangular configuration once an exchange along one of the peripheral red interactions is possible. However, this amount of energy also enables exchanges along the blue interactions --- resulting in a transition between two square/rectangular configurations. A system initialized at the rectangle configuration for $\epsilon \!>\! \epsilon_*$ will rearrange into other rectangular configurations at energies lower than the one required to rearrange to a square geometry. 

To test this hypothesis systematically, we use the Nudge Elastic Band algorithm again, this time setting both the initial and final configurations to be geometrically identical --- the only difference between these is a particle switch along one of the blue interactions. We plot the square-square, rectangle-rectangle energy barriers, $\Delta U_{\rm{S-S}}$ and $\Delta U_{\rm{R-R}}$ respectively, together with the energy at the square-rectangle saddle $U_{\rm{T}}$ shown in Fig.2(a), in Fig.~\ref{fig:FigS1}. The results confirm our hypotheses --- $\Delta U_{\rm{S-S}}$ is comparable to $U_{\rm{T}}$ for $\epsilon \!<\! \epsilon_*$, and $\Delta U_{\rm{R-R}}\!<\!U_{\rm{T}}$ for $\epsilon \!>\! \epsilon_*$. This observation explains the additional plateaus observed for $\epsilon\!>\!\epsilon_*$ for the rectangular system in Fig.3(b). The initial increases in $\lambda_{\rm{max}}$ in Fig.3(b) result from exchanges along the blue interactions that are enabled at lower energies. 

\subsection{Dynamical behaviors}

Figure 5 in the manuscript exemplifies a single trajectory initialized at the local energetic minima, provided with sufficient energy to cross the energetic barrier. We want to show additional dynamics available with the \emph{same} $\epsilon$ and $E_*$ values but with different initial velocities. Figure ~\ref{fig:FigS2} demonstrates several different scenarios.

The top panels show the potential end kinetic energies $U$ and $K$ in (a), the power spectrum of $U$ in (b), and the $D^2_{\rm{min}}$ measure in (c) for a single trajectory. While reminiscent of the trajectory shown in Fig.5, the early-time power spectrum hints that nonlinear effects are present, possibly foreshadowing the upcoming configurational rearrangement.

The middle panels show the same quantities for another trajectory that passes over the barrier quickly --- $U$ and $K$ cross very early on, as shown in panel (d). The early- and late-time power spectra shown in (e) are similar, and the $D^2_{\rm{min}}$ measure in (f) shows the motion has non-negligible non-affine contributions.

Finally, panels (g)-(i) show a particular case where the system does not pass over the energetic barrier (even though it has sufficient energy to do so) and exhibits regular dynamics throughout the entire simulation time, suggesting the trajectory is situated in a stable point in phase-space~\cite{Slichtenberg2013regular}. 

Overall, passing over the barrier and ``escaping'' from the square energy basin seems to be associated with non-negligible, non-affine motion as indicated by the large variations in $D^2_{\rm{min}}$. The emerging ``escape'' time is dramatically affected by the initial conditions. Altogether, the available phase space for specific $\epsilon$ and sufficient $E_*$ to cross the energetic barrier enables the various behaviors exemplified in Fig.~\ref{fig:FigS2}. 



\end{document}